\documentclass[pre,twocolumn,showpacs,superscriptaddress]{revtex4}
\usepackage{amssymb,graphicx}
\begin{document}

\title{A possible generalized form of Jarzynski equality}
\author{Z. C. Tu}
\affiliation{Department of Physics, Tamkang University, Tamsui
25137, Taiwan}
\author{Zicong Zhou}\email{zzhou@mail.tku.edu.tw}
\affiliation{Department of Physics, Tamkang University, Tamsui
25137, Taiwan} \affiliation{Physics Division, National Center for
Theoretical Sciences at Taipei, National Taiwan University, Taipei
10617, Taiwan} 

\begin{abstract}
The crucial condition in the derivation of the Jarzynski equality
(JE) from the fluctuation theorem is that the time integral of the
phase space contraction factor can be exactly expressed as the
entropy production resulting from the heat absorbed by the system
from the thermal bath. For the system violating this condition, a
more general form of JE may exist. This existence is verified by
three \textit{Gedanken} experiments and numerical simulations, and
may be confirmed by the real experiment in the nanoscale.
\end{abstract}\maketitle

\section{Introduction}
Consider a classic system in contact with a thermal bath at
constant temperature, and at some time interval, the system is
driven out of the equilibrium by an external field. Two groups of
equalities are proved to still hold for this system. One is the
fluctuation theorem (FT)
\cite{Evansprl93,Cohenprl95,Evansap02,Wangprl02,Seifert04} which
reflects the probability of violating the Second Law of
Thermodynamics in the non-equilibrium process. Another is the
Jarzynski equality (JE)
\cite{Jarzynskiprl97,Crooksjsp98,Adibpre05} which ensures us to
extract the free energy difference between two equilibrium states
from the non-equilibrium work performed on the system in the
process between these two states. The quantum versions
\cite{Yukawajpsj,Mukamelprl03,Roeckpre04} and experimental
verifications \cite{Hummerpnas,Liphardtsci02} of JE are also
presented. After it was proposed in 1997, the JE has aroused some
controversy
\cite{Cohenjsm04,Gross05,Sung0506,Jarzynskijsm,Luajpcb,Benaepl,Biercm05,Silbey},
in which two typical \textit{gedanken} experiments are quite
interesting and we summarize them as follows.

\begin{figure}[!htp]\begin{center}
\includegraphics[width=6.8cm]{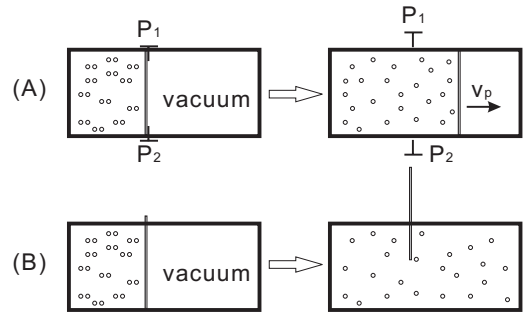}
\caption{\label{idealgas} \textit{Gedanken}
experiments.}\end{center}\end{figure}

Experiment 1: As shown in Fig.~\ref{idealgas}A, imagine that a
closed container, in contact with a thermal bath at constant
temperature, is divided into two compartments by a perfectly thin,
frictionless but heavy enough piston, and imagine that one
compartment initially contains ideal gas of $N$ (large enough)
particles in equilibrium at temperature $T$, while another
compartment is empty. At time $t_1$, we remove the pins $P_1$ and
$P_2$, and give the piston a large initial velocity $v_p$. The gas
will fill the whole container with the movement of the piston.
After a long time relaxation, the system arrives at an equilibrium
state at time $t_2$.

Experiment 2: As shown in Fig.~\ref{idealgas}B, imagine that a
closed container, in contact with a thermal bath at constant
temperature, is divided into two compartments by a perfectly thin
and frictionless plate, and imagine that one compartment initially
contains ideal gas of $N$ particles in equilibrium at temperature
$T$, while another compartment is empty. At time $t_1$, we pull up
the plate, and the gas will expand and fill the whole container.
After a long time relaxation, the system reaches an equilibrium
state at time $t_2$. Here $t_1$ and $t_2$ do not require to have
the same values as those in experiment 1.

Assume the initial volume of the gas to be $V_1$, the whole volume
of the container to be $V_2$. There are two common points in the
above two experiments: (i) The macroscopic work in the expansion
process is vanishing (i.e. $W=0$); (ii) After the systems arriving
at the final equilibrium states, the free energy difference is
$\Delta F=-NT\ln (V_2/V_1)$. The important difference between them
is that the microscopic work, $w$, in the first experiment is
non-vanishing although the macroscopic one $W=\langle w\rangle=0$
for $v_p\rightarrow\infty$ \cite{Luajpcb}, while $w=0$ for the
second one. Due to this difference, the JE holds in the first
experiment (i.e. $\langle e^{-\beta w}\rangle=e^{-\beta \Delta
F}=V_2^N/V_1^N$ with $\beta=1/T$) \cite{Luajpcb} but fails in the
second one ($\langle e^{-\beta w}\rangle=1, e^{-\beta \Delta
F}=V_2^N/V_1^N$). Jarzynski and Crooks argued that the JE fails
because the initial distribution function is not canonical in the
second case \cite{Jarzynskijsm,Crooksthesis}. We would like to
consider this problem from another point of view: the initial
distribution function is still canonical but a more underlying
reason makes the JE fail. In other words, there is a more general
form of JE. The rest of this paper is focus on this topic and
organized as follows: In Sec.~\ref{derivJE}, we sketch the
derivation of JE from the FT and emphasize that the condition of
adiabatic incompressibility \cite{Evansbook} is crucial to this
derivation. In fact, Jarzynski's original proof
\cite{Jarzynskiprl97,Jarzynskijsm} also requires this condition.
In Sec.~\ref{secGJE}, we check whether this condition holds or not
in the above two \textit{Gedanken} experiments and put forward a
generalized JE, Eq.(\ref{genJE3}). The third \textit{Gedanken}
experiment intermediating between the above two experiments is
proposed and the corresponding numerical simulation verifies the
existence of generalized JE. In Sec.~\ref{conclusion}, we give
further discussions and a brief summary.

\section{Derivation of JE from FT\label{derivJE}}
An important relation between the FT and JE is that the JE can be
derived from the FT for time reversible stochastic or
deterministic dynamics \cite{Crookspre99,Evansmp03}. Here we look
through the main idea of Evans' derivation \cite{Evansmp03}. The
phase space of the $N$ particle system is denoted by
$\{\mathbf{q};\mathbf{p}\}$, where $\mathbf{q}\equiv
\{q_{1x},q_{1y},q_{1z},\cdots, q_{Nx},q_{Ny},q_{Nz}\}$ and
$\mathbf{p}\equiv \{p_{1x},p_{1y},p_{1z},\cdots,
p_{Nx},p_{Ny},p_{Nz}\}$ represent the configuration and momentum
spaces, respectively. The phase space contraction factor,
$\Lambda=\frac{\partial \dot{\mathbf{q}}}{\partial
\mathbf{q}}+\frac{\partial \dot{\mathbf{p}}}{\partial
\mathbf{p}}$, depends on the detail dynamics of the system.

Assume that the classic system contacts with a thermal bath at
constant temperature $T$, and that it stays at an equilibrium
state for time $t\le t_1$. Take a microscopic state $A_1$
corresponding to this equilibrium state. From time $t_1$ to
$t'_2$, we switch on an external field denoted by a parameter
$\lambda$ varying from $\lambda_1$ to $\lambda_2$, and drive the
system out of equilibrium. After a sufficient relaxation with
fixed $\lambda_2$, the system arrives at the other equilibrium
state at time $t_2$. Correspondingly, the microscopic state
evolves to $A_2$. From time $t_1$ to $t_2$, the entropy production
function along the microscopic path $\gamma(t)$ linking the states
$A_1$ and $A_2$ is expressed as \cite{Evansap02}
\begin{equation}\label{entropfun1}
s[\gamma(t)]=\ln(f_1/f_2)-\int_{t_1}^{t_2} \Lambda[\gamma(t)] dt,
\end{equation}
where $f_1$ and $f_2$ are the equilibrium distribution functions
at time $t_1$ and $t_2$, respectively. One can prove the FT, $
{p_F(s)}/{p_R(-s)}=e^{s}$, where $p_F(s)$ and $p_R(s)$ represent
the probability distributions of the entropy production function
taking value $s$ along the microscopic path $\gamma(t)$ and its
time-reversal path, respectively. If averaging $e^{-s}$ for all
paths beginning from all microscopic states corresponding to the
macroscopic equilibrium state at time $t_1$, we have
\begin{equation}
\langle e^{-s}\rangle=\int e^{-s}p_F(s)d s=\int p_R(-s)d
s=1.\label{Hat-saseq}
\end{equation}
This is nothing but the Kawasaki identity or Hatano-Sasa equality
\cite{Kawasaki67,HatanoSasa}.

If taking canonical distributions for the initial state at time
$t_1$ and the final state at time $t_2$, we have
$f_1=e^{\beta(F_1-H_1)}$ and $f_2=e^{\beta(F_2-H_2)}$, where $F_1$
and $F_2$ are the free energies of the system at time $t_1$ and
$t_2$ while $H_1$ and $H_2$ are the Hamiltonians of the system at
time $t_1$ and $t_2$. Assume that the effective dynamics of the
system can be expressed as \cite{Evansbook}
\begin{eqnarray}
\dot{\mathbf{q}}_n&=&\partial H /\partial \mathbf{p}_n,\label{dyn1-1}\\
\dot{\mathbf{p}}_n&=&-\partial H /\partial \mathbf{q}_n-
\alpha[\gamma(t)]
 \mathbf{p}_n ,\label{dyn1-2}
\end{eqnarray}
where $\alpha[\gamma(t)]$ is the thermostat multiplier
\cite{Evansbook} ensuring the kinetic temperature of the system to
be fixed at $T$, and it reflects the heat exchange between the
system and the thermal bath. $H$ is the $\lambda$-dependent
Hamiltonian. Under the above dynamics, the phase space contraction
factor is derived as $\Lambda [\gamma(t)]=- 3N \alpha[\gamma(t)]$
and its integral from time $t_1$ to $t_2$ is just the entropy
production induced by the heat ($q[\gamma(t)]$) absorbed by the
system from the thermal bath along the microscopic path
$\gamma(t)$ linking the states $A_1$ and $A_2$, i.e.,
\begin{equation}
\int_{t_1}^{t_2}\Lambda[\gamma(t)] dt=\beta
q[\gamma(t)].\label{crucialeq}\end{equation} This equation is
crucial to the derivation of JE from the FT. Thus the entropy
production function, Eq.~(\ref{entropfun1}), is transformed into
\begin{equation}
s[\gamma(t)]=\beta(w[\gamma(t)]-\Delta F),\label{tempr1}
\end{equation}
where $w[\gamma(t)]=H_2-H_1-q[\gamma(t)]$ (microscopic energy
conservation) is the work performed on the system along the
microscopic path $\gamma(t)$. $\Delta F=F_2-F_1$ is the free
energy change of the system from time $t_1$ to $t_2$. Assume that
$w[\gamma(t)]$ takes value $w$ when $s[\gamma(t)]$ has value $s$,
and notice that there is no work from time $t'_2$ to $t_2$ because
the parameter $\lambda$ is unchanged at this time interval. From
Eqs.~(\ref{Hat-saseq}) and (\ref{tempr1}) we easily arrive at the
JE,
\begin{equation}\langle e^{-\beta w}\rangle=e^{-\beta \Delta
F}.\label{je1997}\end{equation}

We emphasize again that Eq.~(\ref{crucialeq}), the time integral
of the phase space contraction factor exactly expressed as the
entropy production resulting from the heat absorbed by the system
from the thermal bath, is the crucial point in the derivation of
the JE from the FT. Remember that the phase space contraction
factor depends on the microscopic dynamics. If the dynamics
satisfies the condition of adiabatic incompressibility
\cite{Evansbook}, i.e., the phase space contraction factor depends
merely on the thermostat multiplier, Eq.~(\ref{crucialeq}) holds
and the JE is a natural corollary of the FT.

\section{Generalized JE\label{secGJE}}

Now, we check whether the condition of adiabatic incompressibility
holds for the systems mentioned in the above two experiments.

For the first experiment, because the initial velocity
distribution obeys the Maxwell distribution, some particles with
velocity larger than $v_p$ will strike the piston and then bounce
fully but $v_p$ is unchanged because the mass of the piston is
much larger than the total mass of the particles that collide with
it. In each bounces, the piston will do a small work on the gas
system. Lua \emph{et al.} have proved that the mean work is
vanishing but the JE still holds for $v_p\rightarrow\infty$
\cite{Luajpcb}. The effective dynamics can be expressed as
\cite{Evansbook}
\begin{eqnarray}
\dot{\mathbf{q}}_n&=&\mathbf{p}_n/m+\dot{\mathbf{R}}\cdot\mathbf{q}_n,\label{dyn2-1}\\
\dot{\mathbf{p}}_n&=&-\dot{\mathbf{R}}\cdot\mathbf{p}_n-
\alpha[\gamma(t)]
 \mathbf{p}_n ,\label{dyn2-2}\\
 \dot{V}&=&V\dot{r},\label{dyn2-3}
\end{eqnarray}
where $m$ is the mass of each particle, the matrix
$\dot{\mathbf{R}}=\left((\dot{r},0,0)^t,
(0,0,0)^t,(0,0,0)^t\right)$, and $\dot{r}$ the time-dependent
volume expansion ratio with vanishing value except at the time
interval between $t_1$ and $t_2'$. Here the volume $V$ plays the
role of the parameter $\lambda$ in Evans' derivation of JE from
the FT. The phase space contraction factor is found to be $\Lambda
[\gamma(t)]=-3N \alpha[\gamma(t)]$. Hence Eq.~(\ref{crucialeq}) as
well as the condition of adiabatic incompressibility still holds
and so the JE is valid in this experiment.

The effective term $-\dot{\mathbf{R}}\cdot\mathbf{p}_n$ in
Eq.~(\ref{dyn2-2}) reflects the collisions between the piston and
particles. For the second experiment, the volume expansion has no
direct effect on the momentum of the particles. Thus the term
$-\dot{\mathbf{R}}\cdot\mathbf{p}_n$ should be removed, and
Eq.~(\ref{dyn2-2}) should be replaced by
\begin{equation}\dot{\mathbf{p}}_n=-\alpha[\gamma(t)]
 \mathbf{p}_n,\label{dyn3-2}\end{equation} but Eqs.~(\ref{dyn2-1}) and
(\ref{dyn2-3}) are kept intact in the effective dynamics of the
second experiment. Consequently, we obtain the phase space
contraction factor $\Lambda [\gamma(t)]=N\dot{r}- 3
N\alpha[\gamma(t)]$ and its integral from time $t_1$ to $t_2$
\begin{equation}
\int_{t_1}^{t_2}\Lambda [\gamma(t)]dt=N\ln(V_2/V_1)+\beta q
[\gamma(t)],\label{crucialeq2}
\end{equation}
where we have used $\int_{t_1}^{t_2}
\dot{r}dt=\int_{V_1}^{V_2}d(\ln V)=\ln(V_2/V_1)$ and
$-\int_{t_1}^{t_2}3N \alpha[\gamma(t)]dt=\beta q [\gamma(t)]$.
Obviously, Eq.~(\ref{crucialeq}) as well as the condition of
adiabatic incompressibility does not hold in this experiment and
so the JE fails. However, following the derivation from
Eq.~(\ref{crucialeq}) to Eq.~(\ref{je1997}), and replacing
Eq.~(\ref{crucialeq}) by Eq.~(\ref{crucialeq2}), we obtain a
generalized equality beyond JE:
\begin{equation}
\langle e^{-\beta [w-NT\ln(V_2/V_1)]}\rangle=e^{-\beta \Delta
F}.\label{genJE1}
\end{equation}
Because $w=0$ and $\Delta F=-NT\ln(V_2/V_1)$ in the second
experiment, the above equation holds although the original form of
JE fails.

Through the above discussions, we know that the JE holds in the
first experiment but fails in the second one. The underlying reason
is that these two experiments have different microscopic dynamics:
One satisfies the condition of adiabatic incompressibility but
another does not. Especially, the second experiment suggests that a
more general form of JE should exist, and we would like to consider
this possibility. Enlightened by Eq.~(\ref{crucialeq2}), we divide
the time integral of the phase space contraction factor into two
parts: One is $\beta q [\gamma(t)]$, the entropy production
resulting from the heat absorbed by the system from the thermal
bath; Another is the entropy induced by the change of the external
parameter and expressed as $\beta \sigma$. That is,
\begin{equation}
\int_{t_1}^{t_2}\Lambda [\gamma(t)]dt=\beta (q
[\gamma(t)]+\sigma).\label{crucialeq3}
\end{equation}
Following the derivation from Eq.~(\ref{crucialeq}) to
Eq.~(\ref{je1997}) and replacing Eq.~(\ref{crucialeq}) by
Eq.~(\ref{crucialeq3}), we arrive at a generalized JE:
\begin{equation}
\langle e^{-\beta (w-\sigma)}\rangle=e^{-\beta \Delta
F},\label{genJE3}
\end{equation}
which is transformed into \begin{equation} \langle e^{-\beta
w}\rangle=e^{-\beta (\Delta F+\sigma)}\label{genJE33}\end{equation}
if $\sigma$ depends only on the value of the external parameter at
time $t_1$ and $t_2$, but not explicitly on the microscopic pathes.
We conjecture that $\sigma=0$ for most macroscopic system and then
Eq.~(\ref{genJE3}) is degenerated into the original JE. $\sigma\neq
0$ only in some very special systems and correspondingly the
original JE fails. For example, $\sigma=NT\ln(V_2/V_1)\neq 0$ in the
second experiment.

Noticing that the mass of the piston in the first experiment is
infinitely large. If it has an infinitesimal value, this system is
equivalent to that in the second experiment because the collisions
between particles and the piston has no effect on the momentum of
the particles such that the particles do not feel the existence of
the piston. It is interesting to discuss the intermediate case
between the above two limits. Let us consider the third
\textit{gedanken} experiment where the experimental setup is the
same as the first one except the mass $M$ of the piston is finite.
At time $t_1$, we remove the pins $P_1$ and $P_2$, and the gas
will push the piston to the right wall of the container. Once the
piston contacts with the wall, it adheres to the wall without
bounce. After a long time relaxation, the system arrives at an
equilibrium state at time $t_2$. When we write the effective
dynamics, Eq.~(\ref{dyn2-2}) should be replaced by
\begin{equation}\dot{\mathbf{p}}_n=-g\dot{\mathbf{R}}\cdot\mathbf{p}_n-
\alpha[\gamma(t)]
 \mathbf{p}_n ,\label{dyn4-2}\end{equation} but Eqs.~(\ref{dyn2-1}) and
(\ref{dyn2-3}) are unchanged, where $g$ is a function of $m$ and
$M$ taking values between 0 and 1. $g$ may also depend on $N$ and
$V_2/V_1$ because the equations of motion are just the effective
ones. With this dynamics, we obtain $\sigma=(1-g)NT\ln(V_2/V_1)$
from Eq.~(\ref{crucialeq3}). Thus Eq.~(\ref{genJE3}) gives
\begin{equation}\ln \langle e^{-\beta w}\rangle=gN\ln(V_2/V_1).\label{genJE5}\end{equation}

\begin{figure}[!htp]\begin{center}
\includegraphics[width=6.8cm]{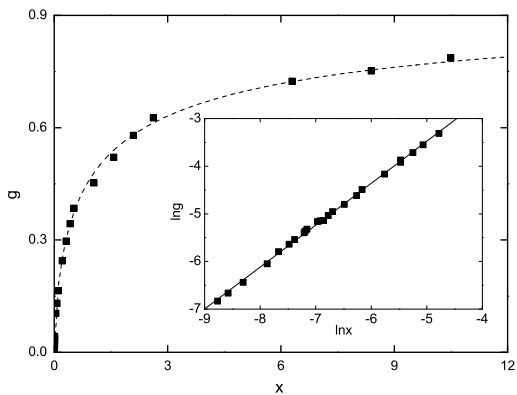}
\caption{\label{gnMcurve} Numerical results and fitting curve for
the relation between $g$ and $x$ where $x$ represents
$M/[mN\ln(V_2/V_1)]$. The squares come from numerical simulations.
The result for small $x<0.01$ is magnified in the inset of the
figure.}\end{center}\end{figure}

In order to recover the former two experiments, $g$ must satisfy
$g\rightarrow 1$ for $M \rightarrow \infty$ and $g\rightarrow 0$
for $M \rightarrow 0$. To determine $g$, we do numerical
simulations for ideal gas with different $N$ (from 1000 to 10000),
$M/m$ (from 0.2 to 1000), and $V_2/V_1$ (from 1.1 to 1.9), and
calculate $g$ by Eq.~(\ref{genJE5}). To obtain the ensemble
average $\langle e^{-\beta w}\rangle$, we take 500 systems
\footnote{We also calculate the ensemble average by using 100
systems and obtain the similar results.} with different initial
microstates corresponding to the same macroscopic equilibrium
state. We find that $g$ depends only on the combined variable
$x=M/[mN\ln (V_2/V_1)]$. The relation between $g$ and $x$ is shown
in Fig.~\ref{gnMcurve}. For very small $x$, the numerical data
(the inset of Fig.~\ref{gnMcurve}) can be fit well by a line $\ln
g=0.93+0.88\ln x$. That is, $g$ has the asymptotic form $g\sim
(2.87x)^{0.88}$ for $x\rightarrow 0$ (corresponding to $M
\rightarrow 0$). Based on this asymptotic form, noting that
$g\rightarrow 1$ for $x \rightarrow \infty$ (corresponding to $M
\rightarrow \infty$), we conjecture that $g$ has the form
\begin{equation}
g=\left[\frac{(2.87x)^{\nu}}{1+(2.87x)^{\nu}}\right]^{0.88/\nu}.\label{fiteqtion}\end{equation}
Our numerical data is indeed fitted well by this form. The fitting
curve is the dash line in Fig.~\ref{gnMcurve} with the parameter
$\nu=0.53$. We use this fitting parameter and
Eq.~(\ref{fiteqtion}) to predict $g=0.8795$ for $M/m=4000$,
$N=1000$ and $V_2/V_1=1.1$, which is quite close to the value
$0.8846$ obtained from the numerical simulations. This fact
implies that our conjecture is reasonable although we cannot
intuitively figure out the physical meaning of the numbers 2.87,
0.88 and $\nu=0.53$ in Eq.~(\ref{fiteqtion}).

\begin{figure}[!htp]\begin{center}
\includegraphics[width=6.8cm]{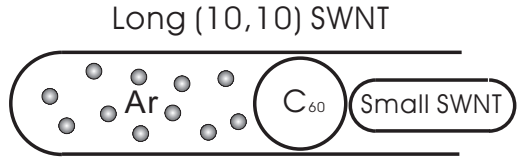}
\caption{\label{nanoexpt} Schematic figure of the experimental
setup (in vacuum).}\end{center}\end{figure}

For the macroscopic gas system except for the case in the second
experiment, we have in general $M\gg mN\ln(V_2/V_1)$, so $g \sim
1$. Hence $\sigma=0$ and Eq.~(\ref{genJE3}) is degenerated into
the JE. Therefore the departure from the JE should occur at the
small scale system with finite $M$ but still large enough $N$. For
example, our result might be verified for the inert gas in a very
long single-walled carbon nanotube (SWNT) as shown in
Fig.~\ref{nanoexpt}. One end of the nanotube is closed while
another is opened, and a buckyball C$_{60}$ is put in it as a
piston. Select the proper nanotube, for example (10,10) nanotube,
and the gas with large radius, for example Ar, such that C$_{60}$
can prevent the gas from escaping from the interstice between
C$_{60}$ and the nanotube. A small SWNT can be used to fix the
initial position of C$_{60}$. At some time, pull the small SWNT
outward to another position quickly. The gas will push C$_{60}$ to
the new position, and one can measure the velocity of C$_{60}$
when it arrives at the new position and calculate the
corresponding work. Repeat this process for many times and
calculate the value of $-\ln\langle e^{-\beta w}\rangle/\beta$.
Comparing this value with the free energy obtained from
theoretical calculation, one can obtain the value of $\sigma$. If
$\sigma\neq 0$, the JE is violated and a generalized JE should
exist.

\section{Discussion and Conclusion\label{conclusion}}

It is useful to discuss some questions before concluding this
paper.

(i) If the volume of the thermal system is fixed but other
parameter varies such as in the single molecule mechanical
experiments \cite{Hummerpnas,Liphardtsci02}, JE always holds
because the effective dynamics can be expressed as
Eqs.(\ref{dyn1-1}) and (\ref{dyn1-2}). Generally speaking, one may
not construct a parameter-dependent Hamiltonian and express the
effective dynamics as Eqs.(\ref{dyn1-1}) and (\ref{dyn1-2}) if the
volume changes. JE may be violated in this case. However, JE still
holds if one can control the ratio of volume change because the
the particles fully bounce when they collide with the piston in
this case, i.e., controlling the ratio is equivalent to
$M\rightarrow \infty$.

(ii) The derivation of JE from FT does not require the thermostat
for the whole process from $t_1$ to $t_2$. For $t\le t_1$, the
system is at an equilibrium state in contact with a thermal bath
at temperature $T$. The external field is switched on from time
$t_1$ to $t_2'$. The contact with a thermal bath is unnecessary in
this stage (i.e. no heat exchange, $q=0$). After that, the
external field is fixed and let the system contact with the same
thermal bath. At time $t_2$, the system reaches the equilibrium
state through a long time relaxation. The thermostat is merely
required in this stage (from $t_2'$ to $t_2$). In fact, this
requirement is the same as the original proof of JE
\cite{Jarzynskiprl97}, which ensures us to calculate $\langle
e^{-\beta w}\rangle$ easily from numerical simulations.

(iii) Equations (\ref{dyn2-1})--(\ref{dyn2-3}), (\ref{dyn3-2}),
and (\ref{dyn4-2}) are the \emph{effective} dynamics of thermal
system with the volume changes. Here the ``effective'' means that
the dynamics is not the real microscopic motion (of course, the
real motion for each particle still abides by Newtonian laws),
while it is the image mapping from the real motion and can give
the correct thermodynamic properties of the system through
Molecule Dynamics Simulations \cite{Evansbook}. The effective term
$-\dot{\mathbf{R}}\cdot\mathbf{p}_n$ in these equations reflects
the collisions between the piston and particles. If the particle
fully bounces (i.e. $M\rightarrow\infty$), the coefficient before
this term is $1$. If the particle do not bounce (i.e.
$M\rightarrow 0$), the coefficient is $0$. For finite $M$, the
coefficient should intermediate between $0$ and $1$. We use $g$ to
express this coefficient in Eq.(\ref{dyn4-2}). In the second
experiment, $V=V_1$ for $t\le t_1$, but $V=V_2$ for
$t=t_2'\rightarrow t_1^{+}$. Thus $\dot{r}$ is a
$\delta$-function, which implies $\mathbf{q}_n$ changes
discontinuously. This is impossible for real dynamics but
permitted in the effective dynamics. Our numerical simulation is
performed for real dynamics (Newtonian mechanics) and the results
reveal that $g\rightarrow 1, \sigma\rightarrow 0$ thus JE holds
for $M\rightarrow \infty$, and that $g\rightarrow
0\Rightarrow\sigma=NT\ln(V_2/V_1)$ thus Eq. (\ref{genJE1}) holds
for $M\rightarrow 0$. That is, the numerical results obtained from
the real dynamics are the same as those derived from the effective
dynamics Eqs. (\ref{dyn2-1})--(\ref{dyn2-3}) and (\ref{dyn3-2}),
which suggests that the effective dynamics is consistent with the
real dynamics and our argument that the term
$-\dot{\mathbf{R}}\cdot\mathbf{p}_n$ reflects the collisions
between the piston and particles is reasonable.

In summary, we have pointed out that the crucial point in the
derivation of the JE from the FT is that the time integral of the
phase space contraction factor is exactly expressed as the entropy
production resulting from the heat absorbed by the system from the
thermal bath, i.e. the dynamics of the system satisfies the
condition of adiabatic incompressibility. For the system violating
this condition, a more general version of JE, Eq.~(\ref{genJE3}),
exists. In the future, it is interesting to find some real systems
which makes $\sigma\neq 0$. Deriving the analytic expression of
the quantity $\sigma$ is another challenge because $\sigma$ might
be system-dependent. These researches will enhance our
understanding to non-equilibrium statistics.

\section*{Acknowledgements}
We are grateful to the useful comments from Prof. C. Jarzynski, U.
Seifert, M. Bier and Dr. Gomez-Marin. This work is supported by
the National Science Council of Republic of China under grant no.
NSC 94-2119-M-032-010 and NSC 94-2816-M-032-003.

\end{document}